\documentclass[nofootinbib,aps, twocolumn]{revtex4}
\usepackage{amssymb,amsmath,graphicx,color,microtype}

\newcommand{\nc}{\newcommand}
\nc{\ba}{\begin{eqnarray}}
\nc{\ea}{\end{eqnarray}}

\begin{document}
\title{Scalar-Tensor Anti-Cross-Correlation does not Help to Reconcile the Tension
 \\ between BICEP2 and Planck}

\author{Razieh Emami $^1$}
\email{email-AT-ipm.ir}

\author{Hassan Firouzjahi$^2$}
\email{firouz-AT-ipm.ir}

\author{Yi Wang $^3$}
\email{yw366-AT-cam.ac.uk}

\affiliation{$^1$School of Physics, Institute for Research in
Fundamental Sciences (IPM),
P.~O.~Box 19395-5531,
Tehran, Iran}

\affiliation{$^2$School of Astronomy, Institute for Research in
Fundamental Sciences (IPM),
P.~O.~Box 19395-5531,
Tehran, Iran}

\affiliation{Centre for Theoretical Cosmology, DAMTP, University of Cambridge, Cambridge CB3 0WA, UK}

\begin{abstract}
  We examine the recent proposal \cite{Contaldi:2014zua} that the anti-cross-correlation of the scalar and tensor primordial fluctuations reconciles the tension between BICEP2 and Planck observations.
 We show that unfortunately the contribution from the scalar-tensor correlation to the CMB temperature power spectrum vanishes once summed over the angular mode $m$ so the tension between BICEP2 and Planck  can not be reconciled. The reason is that one has to use the spin-weighted spherical harmonics when projecting fields with nonzero spins into the CMB sky. These nonzero spin-weighted spherical harmonics are subject to  orthogonality conditions when summed over the angular direction and thus the desired cross-correlation between the scalar and tensor perturbations  in CMB map
vanishes. 

\end{abstract}

\maketitle

The recent discovery of CMB B-mode polarization by BICEP2 \cite{Ade:2014xna} has profound implications for fundamental physics. On the one hand, the B-mode polarization implies  a large amplitude of graviton fluctuations, which consequentially fixes the inflationary energy scale and locally fixes the inflationary potential and its derivatives \cite{Ma:2014vua, Choudhury:2014kma}.  On the other hand, the tensor impact on the B-mode polarization does not seem to be consistent with the tensor impact on the temperature map.

The CMB temperature map has been extensively measured, leaving errors dominated by the cosmic variance, at the range of scales where primordial tensor fluctuations could contribute to the temperature correlations. From the temperature map, the WMAP + ACT + SPT + BAO + $H_0$ constraint of $r$ is $r < 0.13$ at $95\%$ CL \cite{Hinshaw:2012aka}. And the Planck constraint is $r < 0.11$ at $95\%$ CL \cite{Ade:2013uln}. The tension between those measurements and the BICEP2 detection $r=0.20^{+0.07}_{-0.05}$ is about $3 \sigma$.
 \cite{Smith:2014kka}.

As an intriguing proposal to reconcile the tension between Planck and BICEP2 observations, the authors in \cite{Contaldi:2014zua}  suggested that  when the inflationary fluctuations are statistically anisotropic, there can be cross-correlations between the tensor mode(s) and the scalar mode. Such a cross-correlation, if negative, may help to suppress the inflationary power spectrum at low multiple moment $l$. Note  that the tension between Planck and BICEP2 mainly comes from the fact that  Planck observes a suppression of power at low $l$ at $5\%\sim 10\%$ (assuming no tensor mode) while BICEP2 observes tensor mode which should have enhanced the low $l$ power spectrum by 
$10\%$. Based on such a mismatch at about $15\% \sim 20\%$, it is hoped that the anti-correlation between the scalar and tensor perturbations can lower the small $l$ temperature power and thus reconcile the tension. 

In this short note we show that unfortunately such a suppression is not possible. This is because we have to use  the spin-weighted spherical harmonics when projecting fields with nonzero spins in the CMB map. These nonzero-spin-weighted spherical harmonics have 
certain orthogonality  relations when summed over the angular direction and thus the CMB cross-correlation vanishes. 

This conclusion was specifically demonstrated  in our earlier paper \cite{Chen:2014eua} for the 
models of anisotropic inflation in which the background rotational symmetry is broken 
in the presence of a vector field during inflation  \cite{Emami:2010rm}.  Here we present the proof 
for the general case and model-independently. 

To see this explicitly, one can expand the CMB temperature fluctuation with primordial perturbations and spherical harmonics (see \cite{Kamionkowski:1996zd, Hu:1997hp, Watanabe:2010bu, Kamionkowski:2014faa} for details). The CMB temperature anisotropy $\delta T$ is related to $a^{T}_{l, m}$ as
\begin{align}
  a^{T}_{l, m} = \int d\Omega  ~ \delta T(\theta, \phi) Y_{lm}(\theta, \phi)~.
\end{align}
Further, expanding $\delta T$ as an integral of radiation transfer function and primordial fluctuations, we have
\ba
a^{T}_{l, m} = 4\pi \sum_{i=-2,0,2} \int  \frac{d^3 k}{(2\pi)^3} \Delta_l^{i T}(k) h_i(\mathbf{k}) [{}_iY^*_{lm}(\theta,\phi)]~,
\ea
where $\Delta_l^{i T}(k)$ is the radiation transfer function transferring primordial component $i$ into temperature. Here $h_i$ collectively denotes primordial fluctuations. We shall restrict our attention to scalar and tensor primordial fluctuations and neglect the typically decaying vector modes. The $h_i$'s can be explicitly written down as
\begin{align}
  h_0 = \zeta , \quad h_{-2} = \frac{1}{\sqrt 2} \left( h^+ + i h^\times \right)~,
  \quad h_{2} = \frac{1}{\sqrt 2} \left( h^+ - i h^\times \right) 
\end{align}
where $\zeta$ is the curvature perturbation while $h^+$ and $h^\times$ represent the two polarization modes of the gravitational waves.

Here ${}_{i}Y_{lm}(\theta, \phi)$ is the spin-$i$-weighted spherical harmonics and we shall need to sum over $i = (0, -2, 2)$ for the scalar and tensor fluctuations respectively. It is important to note that one can not expand a tensor field using the  zero spin-weighted spherical harmonics ${}_0Y_{lm}(\theta,\phi) \equiv Y_{lm}(\theta,\phi)$, because otherwise the transformation properties does not match under rotation.

The two point correlation function of $a^{T}_{l, m}$ is
\ba
\label{eq:aXaX}
\langle a^{T}_{l, m} a^{T}_{l, m}\rangle &=& 4 \pi \sum_{i_1,i_2}
\int \frac{dk}{k} \Delta_{l}^{i_1 T}(k) \Delta_{l}^{i_2 T}(k)    \\
&\times& \int  d\Omega 
  [{}_{i_1}Y^*_{lm}(\theta, \phi)] [{}_{i_2}Y_{lm}(\theta, \phi)] P^{i_1, i_2}(k, \theta, \phi)
  \nonumber
\ea
Here the $P^{i_1, i_2}(k, \theta, \phi)$ is the dimensionless correlation function between the $(i_1, i_2)$ components of perturbations 
\ba
\label{eq:pdef}
  P^{0, \pm 2} & = &  \frac{\left( P^{0, +} \pm i P^{0, \times}\right)}{\sqrt{2} } , ~~
  P^{\pm 2, 0} = \frac{ \left( P^{+, 0} \mp i P^{\times, 0}\right)}{\sqrt{2}}
\nonumber\\
P^{\pm 2,\pm 2}  &=&  \frac{\left( P^{+, +}+P^{\times, \times} \right)}{2}  , ~~
  P^{\pm 2,\mp 2} = \frac{\left( P^{+, +}-P^{\times, \times} \right)}{2}  \nonumber\\
\ea
For example, in anisotropic inflation, $P^{0,+}$ is non-vanishing and can indeed be negative \cite{Watanabe:2010bu, Chen:2014eua}. This is because in anisotropic inflation, the three dimensional rotational symmetry is broken. The conventional scalar, vector and tensor fluctuations during inflation thus no longer belong to different irreducible representations of the $SO(3)$ rotation group. As a result, cross-correlations between these modes become possible at the level of linear perturbation theory. Anisotropic inflation also predicts $P^{0, \times}=0$, and the difference in $P^{+, +}$ and $P^{\times, \times}$ is suppressed by the smallness of the anisotropy. However, here our discussion shall not depend on explicit realization of anisotropies and we leave \eqref{eq:pdef} to be general and all components can be turned on and/or being unequal.

When studying anisotropies, one does not need to sum over the angular index $m$. However, when addressing the tension issue between BICEP2 and Planck, $m$ has to be summed over because the $m$-averaged observable
\begin{align}
  C_l^{TT} = \frac{1}{2l+1} \sum_{m=-l}^l \langle a^{T}_{l, m} a^{T}_{l, m}\rangle
\end{align}
is the quantity that has been addressed in the tension problem. By summing over $m$, only the ($i_1=0$, $i_2=0$) and ($i_1=2$, $i_2=2$) pairs survive into $C_l^{TT}$. The rest parts vanish. For example, consider ($i_1=-2$, $i_2=0$). We have
\begin{align}
  C_l^{TT, -2,0} = \int d\Omega \sum_{m=-l}^l [{}_{-2}Y^*_{lm}(\theta, \phi)] [{}_{0}Y_{lm}(\theta, \phi)] \times \cdots,
\end{align}
where $\cdots$ denote terms that do not depend on $m$. The summation can be calculated as \cite{Kamionkowski:1996zd,Hu:1997hp}
\begin{align}
  \sum_{m=-l}^l [{}_{-2}Y^*_{lm}(\theta, \phi)] [{}_{0}Y_{lm}(\theta, \phi)] = \sqrt{\frac{2l+1}{4\pi} } Y_{l,2}(0,0) = 0~,
\end{align}
where the relation $Y_{l,2}(0,0)=0$ can be shown by expanding the spherical harmonic function into the associated Legendre polynomial and note that $P^M_L(\cos\theta) \propto (\sin^2\theta)^{M/2}$. For calculating $Y_{l,2}(0,0)$, $M=2$ and thus we end up with $Y_{l,2}(0,0)\propto \sin^2\theta = 0$.

Similarly, for the ($i_1=2$, $i_2=0$) case we have 
\begin{align}
  C_l^{TT, 2,0} \propto  Y_{l,-2}(0,0) = Y^*_{l,2}(0,0) = 0~.
\end{align}
Finally the ($i_1=0$, $i_2=2$) and ($i_1=0$, $i_2=-2$) cases are complex conjugate of the above considered cases which also vanish.
As a result, even with the presence of anisotropy, $\langle\zeta h_{\pm2}\rangle$ cannot contribute to the $C_l^{TT}$ correlation.

It is also worth mentioning that although the temperature power spectrum $C_l^{TT}$ does not receive contribution from scalar-tensor cross-correlation, there are contributions from $\langle h_{\pm 2}h_{\pm2}\rangle$, and also possibly from an anisotropic part of the $\langle\zeta \zeta\rangle$ correlation. However, those parts do not suppress the low $l$ temperature spectrum either. For the $\langle h_{\pm 2}h_{\pm2}\rangle$ part, this is because the correction is positive definite, as one can see from equation \eqref{eq:aXaX}. On the other hand, the anisotropic part of the $\langle\zeta \zeta\rangle$, if exists and has similar scale dependence as the isotropic part, gives an overall rescaling of the temperature power spectrum. The shape of the spectrum is not modified, and the amplitude modification is subject to re-normalizing the power spectrum which is irrelevant to the BICEP2-Planck tension. In addition, it has been proved recently in \cite{Kamionkowski:2014faa} that the contribution from $\langle h_{\pm 2}h_{\mp2}\rangle$ under the Copernican principle, here the homogeneity of space, is also equal to zero.

Finally, we would like to mention that the conclusion above does not provide any no-go theorem
for studying anisotropies in general. Because if we are not to address the tension issue, we are free to study the TT, TE, EE, TB, EB, BB correlators without averaging over $m$, and also not only the $l_1=l_2$ ones but also the $l_1=l_2\pm 1$ and $l_1=l_2\pm 2$ ones. Thus there is rich phenomenology to study after the gravitational wave detection  \cite{Chen:2014eua}.

With these discussions in mind, the primordial anisotropies in gravitational waves are curious effects. 
There may be a deep connections between anisotropies in gravitational waves and the low-$l$ CMB anomaly as advocated recently in \cite{Chen:2014eua} which deserves further investigations.
In addition, as argued in  and \cite{Dai:2013kfa},  \cite{Abolhasani:2013vaa} and 
\cite{Chluba:2014uba}, the anisotropies in gravitational waves may also be connected to  hemispherical asymmetry as observed by Planck. These possibilities  make the search for the features in the B-mode polarization, such as statistical anisotropies, more interesting.  \\

Note added: While this work was completed the paper \cite{Zibin} appeared which also reached the same conclusion as in this paper. Note that the conclusion that the tensor-scalar anti-correlation does not help to reconcile the tension between the Planck and BICEP2 observations were explicitly demonstrated in the context of anisotropic inflation in our earlier paper \cite{Chen:2014eua}.
Here we demonstrated it  in generality. 

\bigskip
\noindent \textit{Acknowledgments.}
The authors are grateful to X. Chen  and S.-C. Su for the insightful discussions. 
We also thank X. Chen for the collaboration in a related work \cite{Chen:2014eua} and his comments about the current draft. YW is supported by a Starting Grant of the European Research Council (ERC STG grant 279617), and the Stephen Hawking Advanced Fellowship.

\end{document}